\documentclass{aastex63}

\usepackage{hyperref}

\definecolor{mygreen}{rgb}{0.0, 0.5, 0.0}
\definecolor{mylila}{rgb}{0.5, 0.0, 0.5}
\definecolor{myblue}{rgb}{0.0, 0.0, 0.5}
\definecolor{mypurple}{rgb}{0.58, 0.42, 0.87}

\begin{document}

\title{TESS Input Catalog versions 8.1 and 8.2: Phantoms in the 8.0 Catalog and How to Handle Them}


\author[0000-0001-8120-7457]{Martin Paegert}
\affiliation{Center for Astrophysics $\vert$ Harvard \& Smithsonian, 60 Garden Street, Cambridge, MA 02138, USA}

\author[0000-0002-3481-9052]{Keivan G. Stassun}
\affil{Vanderbilt University, Department of Physics \& Astronomy, 6301 Stevenson Center Ln., Nashville, TN 37235, USA}

\author[0000-0001-6588-9574]{Karen A.\ Collins}
\affiliation{Center for Astrophysics $\vert$ Harvard \& Smithsonian, 60 Garden Street, Cambridge, MA 02138, USA}

\author[0000-0002-3827-8417]{Joshua Pepper}
\affiliation{Department of Physics, Lehigh University, 16 Memorial Drive East, Bethlehem, PA 18015, USA}

\author[0000-0002-5286-0251]{Guillermo Torres}
\affiliation{Center for Astrophysics $\vert$ Harvard \& Smithsonian, 60 Garden Street, Cambridge, MA 02138, USA}

\author[0000-0002-4715-9460]{Jon Jenkins}
\affiliation{NASA Ames Research Center, Moffett Field, CA 94035-0001}

\author[0000-0002-6778-7552]{Joseph D. Twicken}
\affiliation{SETI Institute, Mountain View, CA  94043, USA}
\affiliation{NASA Ames Research Center, Moffett Field, CA  94035, USA}

\author[0000-0001-9911-7388]{David W.\ Latham}
\affiliation{Center for Astrophysics $\vert$ Harvard \& Smithsonian, 60 Garden Street, Cambridge, MA 02138, USA}

\begin{abstract}
We define various types of ``phantom" stars that may appear in the TESS Input Catalog (TIC), and provide examples and lists of currently known cases. We present a methodology that can be used to check for phantoms around any object of interest in the TIC, and we present an approach for correcting the TIC-reported flux contamination factors accordingly. We checked all 2077 TESS Objects of Interest (TOIs) known as of July 21st 2020 (Sectors 1--24) and sent corrections for 291 stars to MAST where they are integrated into the publicly available TIC-8, updating it to TIC 8.1. We used the experience gained to construct an all-sky algorithm searching for ``phantoms" which led to 34 million updates integrated into TIC 8.2.
\end{abstract}

\section{Introduction}\label{sec:intro}

The TESS Input Catalog \citep[TIC;][]{Stassun:2019} is a comprehensive collection of $1.73\times10^9$ sources on the sky, 
providing stellar parameters for evaluation of potential planetary transit signals. It was constructed for the TESS mission to serve as a source for selecting targets to observe with the TESS 2-minute cadence, and to provide stellar parameter information for evaluating the properties of transit candidates.  The TIC was deliberately constructed to include stars much fainter than the set of likely TESS targets, since fainter stars near a transit candidate present two potential types of problems.  A faint eclipsing binary system blended with a brighter target star can produce a photometric signal consistent with a planet-size transit of the bright star.  Also, the light from stars close to a target star can fall into the target star aperture, diluting a real transit of the target star, and therefore affecting the apparent size of the potential planet, a phenomenon referred to as flux contamination.
The TIC was built to include as many faint stars as possible to allow the mission to account for those effects. 

Despite great pains taken to ensure the purity of the TIC, certain types of spurious entries in such a large catalog representing the federation of multiple smaller catalogs are unavoidable. And while representing a small fraction of the overall TIC, left alone they can lead to incorrectly estimated transiting planet radii, which can cause wasted followup effort on otherwise low-priority planets or deprioritization of otherwise mission-critical planets. 

The way in which we can account for these issues in modifications and updates to the TIC is complicated by rules governing the TIC.  In order to ensure that early analysis and publications using TIC designations and information are interpretable by later users, we have adopted certain stipulations in updating the TIC.  Two of the key rules are that no TIC ID is ever deleted, and that the TIC ID associated with a given star is not changed from one version of the TIC to the next.  This is to ensure maximum backwards compatibility when using TIC ID or stellar information. For the vast majority of stars, such rules present no complication, but in cases when our knowledge of the underlying number or existence of stars changes due to new information, we have to construct highly specific flags and designations to note and account for these types of changes. 

As new astronomical surveys are released, especially the release of {\it Gaia\/} DR2 \citep{2016Gaia} between the construction of TIC-7 \citep{Stassun:2018tic} and TIC-8 \citep{Stassun:2019}, we attempted to maintain both completeness and reliability.  In most cases where the available information suggested that there was likely to be a star at a given location, we included that object in the TIC.  The need to preserve completeness was driven by the belief that it was more important to know about the possible presence of astrophysical sources to identify potential false positives among the TESS transit candidates.  After the release of TIC-8, we have identified various cases where the incorporation of new data, mostly from {\it Gaia\/} DR2, into TIC-7, has led to problems with a small number of objects in TIC-8.  We refer to these problems as ``phantoms", which generally refer to objects listed in TIC-8 that are not real, or where the number of objects associated with a given TIC ID or set of IDs is incorrect.

The purpose of this paper is to provide (i) a methodology for identifying phantoms around any particular TIC objects of interest, and (ii) an approach for correcting the TIC-reported flux contamination factors accordingly.  

Section \ref{sec:definition} describes the definition of three classes of phantoms: artifacts, joins and splits. Section \ref{sec:results} describes the results from manually checking $2077$ TOIs and Appendix~\ref{sec:appendix} contains a list of identified phantoms in this search. Section \ref{sec:allsky} describes the algorithm derived using the experience gained with the manual search. Appendix~\ref{sec:code} describes tools for inspecting any object of interest and for correcting its flux-ratio; the tools can be obtained from \url{https://github.com/mpaegert/tic_inspect}.

\section{Definitions} \label{sec:definition}

\begin{figure}[!ht]
    \centering
    \includegraphics[width=0.235\linewidth]{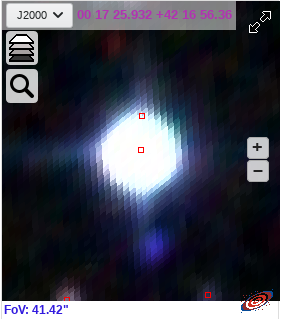}
    \includegraphics[width=0.225\linewidth]{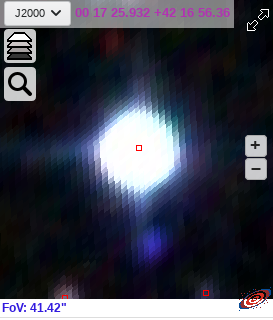}
    \hspace{1cm}
    \includegraphics[width=0.225\linewidth]{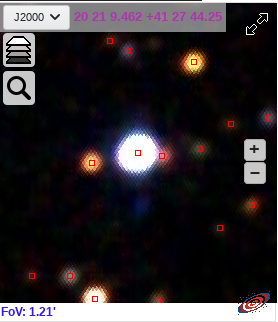}
    \includegraphics[width=0.225\linewidth]{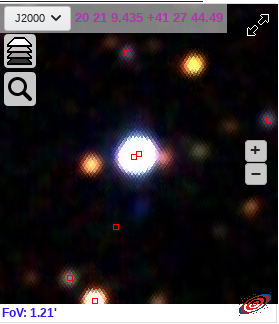}
    \caption{
    {(Left pair:) {\it 2MASS} color images with {\it 2MASS} sources marked (left) and {\it Gaia\/} DR2 targets marked (right). Two {\it 2MASS} sources are represented by a single {\it Gaia} star. The upper {\it 2MASS} detection is located on the northern diffraction spike; it is a {\it 2MASS} Artifact (TIC 440100539).
    }
    (Right pair:)  Same, demonstrating two real {\it Gaia\/} stars represented by a single {\it 2MASS} star. This is an example for a split. Please note that a number of real stars seen by {\it 2MASS} are not seen by {\it Gaia} (The bright reddish star left of the center (TIC 13419950) for example).
    }
    \label{fig:case1}
\end{figure}

For the purposes of this paper, we refer to problem entries in the TIC generally as ``phantoms", which can be caused by one of three situations (visual examples are provided in Figure~\ref{fig:case1} and Figure~\ref{fig:case2}): 

\begin{itemize}
\item Artifact: These are spurious sources from {\it 2MASS} generally caused by diffraction spikes around bright stars. These should be removed from the dilution correction, though being faint the effect will generally be very small. 

\item Join: Two TIC objects of near-equal brightness are in fact the same one real star. They originate from slight mismatches between different catalogs, usually a {\it 2MASS} star that failed to be matched with its respective {\it Gaia} DR2 entry. One star's flux should be removed from the dilution correction. 

\item Split: There are three or more TIC objects, but the brightest one is not real. It is a combination (sum of flux) of two or more fainter ones, which originate from {\it Gaia\/} DR2. The {brightest} star's flux should be removed from the dilution correction. Additional adjustments may be needed depending on the brightness of the true host star. 
\end{itemize}

\section{Results}\label{sec:results}

\subsection{A Methodology for Identifying and Dealing with Phantoms in the TIC}\label{sec:howto}

\subsubsection{Artifacts}\label{sec:artifacts}

\begin{figure}[!ht]
    \centering
    \includegraphics[width=0.225\linewidth]{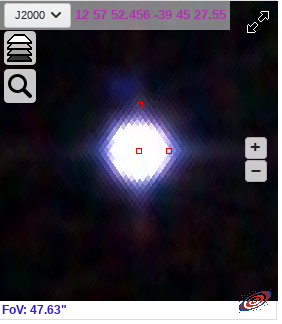}
    \includegraphics[width=0.225\linewidth]{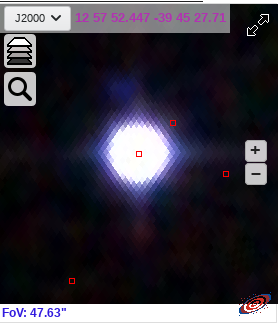}
    \caption{{\it 2MASS} color images for TIC 178819686 with {\it 2MASS} sources marked (left) and {\it Gaia\/} DR2 sources marked (right), showing two Artifacts located on the {\it 2MASS} diffraction spikes. The additional {\it Gaia\/} sources are all fainter than 20th mag in the {\it Gaia\/} $G$ band, and are thus too faint for {\it 2MASS} to detect.}
    \label{fig:artifacts}
\end{figure}

Artifacts originate from {\it 2MASS} and usually sit on the diffraction spikes of bright stars. The effect is most pronounced for red stars ($J > H > K_S$), but we found Artifacts around blue stars as well. TIC-8 includes the {\it 2MASS} quality flags as a dash-separated string (column {\tt TWOMflag}). The first three characters of this string are the photometric quality flags for the {\it 2MASS} bands. They can have the values of {\tt A} to {\tt E} and {\tt U} (best to worst). To identify Artifacts we require two of the flags to be {\tt U} and the star not to be matched to a {\it Gaia\/} DR2 source. Figure~\ref{fig:artifacts} shows a typical case.

We ran a cone-search with a radius of $20\farcs25$ (about 1 TESS pixel) around all 2007 TESS Objects of Interest (TOIs) included in the manual search, and identified 137 potential Artifacts around 121 TOIs which had not been recognized as of July 21 2020 and thus not included in the preliminary tables on the MAST site, and therefore are not yet mentioned in the TIC release notes on MAST\footnote{\url{https://outerspace.stsci.edu/display/TESS/TIC+v8+and+CTL+v8.xx+Data+Release+Notes}}. A sample list of these Artifacts is part of the repository on Github; the full list is integrated into the TIC on MAST. Stars with this disposition should be ignored.

With most of the Artifacts being significantly fainter than their respective nearby TOI (by 3 magnitudes or more), the impact on the calculation of flux contamination for a TOI by more distant Artifacts is minimal. As diffraction spikes of brighter stars are more prominent, the distance from the bright star and the number of Artifacts grows with brightness in the {\it 2MASS} bands.

\subsubsection{Joins}

\begin{figure}[!ht]
    \centering
    \includegraphics[width=0.225\linewidth]{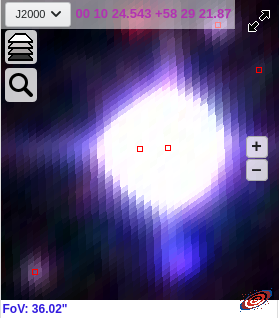}
    \includegraphics[width=0.225\linewidth]{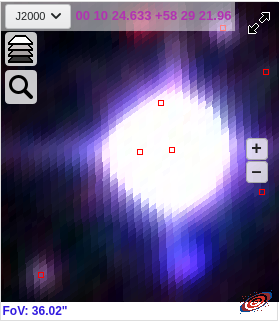}
    \includegraphics[width=0.225\linewidth]{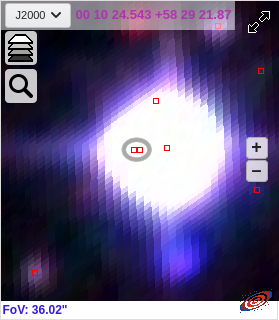}
    \caption{
    (Left:) {\it 2MASS} color images with {\it 2MASS} sources marked.
    (Center:) same image with {\it Gaia\/} sources marked. 
    (Right:) same image with TIC sources marked. Circled TIC sources have been joined as they represent the same star. In this case the {\it 2MASS} $J$ magnitude is of low photometric quality (F) and much too bright which prevented to identify the {\it 2MASS} and {\it Gaia\/} source as being the same star (TIC 396356110).
    }
    \label{fig:case2}
\end{figure}

There are several million stars in TIC-8 that originate from {\it 2MASS} but do not have a match from {\it Gaia\/} DR2. As the right pair of Figure~\ref{fig:case1} illustrates, they are mostly very red stars that have been detected by {\it 2MASS} but not by {\it Gaia}. However, there are cases in which {\it 2MASS} stars do have a previously unnoticed match in {\it Gaia}, as shown in Figure~\ref{fig:case2}. In the left image {\it 2MASS} sources are marked, in the center image {\it Gaia\/} sources are marked, and in the right image sources in TIC-8 are marked. The two circled marks in the right image represent the same star that ended up in TIC-8 twice: once from {\it 2MASS} and once from {\it Gaia}. Both have to be joined in TIC-8. In this particular case all {\it 2MASS} magnitudes are tainted by the brightest star to the right (TIC 396356111) resulting in a TESS magnitude of 7.331, while using {\it Gaia\/} magnitudes results in a TESS magnitude of 11.665. Because of the magnitude difference our automated cross-matching failed to recognize these sources as being the same star. 
Thus we look to see if the target star does not have an ID from {\it Gaia}, and in that case we search for close-by {\it Gaia\/} stars missed in the initial match between {\it 2MASS} and {\it Gaia\/} DR2 in the construction of TIC-8. We employ a default search radius of $5\arcsec$ outside of the Galactic plane \citep[as defined in][]{Stassun:2019} and $4\arcsec$ within the plane. If we find one and only one {\it Gaia\/} counterpart, we deem it to be a candidate for a Join. 

Forty-seven of the 2007 TOIs are stars without a {\it Gaia\/} match; 16 of them are candidates for a Join based on the above criteria. The complete list is part of the repository on Github and MAST. As there are only a few of them, we checked each of them manually before we update the TIC database. In general, we consider the brightness in the TESS band, the quality flags from ({\it 2MASS} and {\it Gaia}), and we inspect the neighbourhood. Let TIC ID~1 be the original entry from {\it 2MASS}, and TIC ID~2 be its potential match from {\it Gaia}. In order to allow existing pipelines to re-process these TOIs, we keep TIC ID~1 as the valid star, but generally copy {\it Gaia\/} properties from TIC ID~2 to TIC ID~1. 

In detail:
\begin{enumerate}
    \item Coordinates are always copied from TIC ID 2 to TIC ID 1.
    \item Proper motions are copied from TIC ID 2 to TIC ID 1 if available from {\it Gaia\/} DR2.
    \item If star 1 is a member of a specially curated list in the TIC \citep[e.g., the Cool Dwarf Catalog; see][]{Stassun:2019}, we simply adopt the stellar parameters from the specially curated list.
    \item If the {\it Gaia\/} DR2 quality flags in TIC-8 indicate that the properties derived from {\it Gaia\/} are acceptable (column {\tt gaiaqflag} is 1) we copy those from TIC ID~2 to TIC ID~1.
    \item Finally we flag TIC ID 2 as DUPLICATE, to prevent it from being taken into account.
\end{enumerate}

\subsubsection{Splits}

\begin{figure}[!ht]
    \centering
    \includegraphics[width=0.225\linewidth]{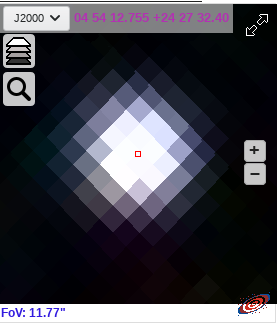}
    \includegraphics[width=0.225\linewidth]{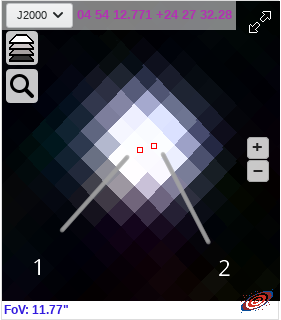}
    \caption{{\it 2MASS} color images, with one {\it 2MASS} source identified on the left, and two {\it Gaia\/} sources identified on the right.  This is the case of a Split, where the single object was TIC 60340880 in TIC-7.  The two {\it Gaia\/} DR2 targets marked are GAIA 3419713335430708352 as star 1, GAIA 3419713335430708480 as star 2, about $0\farcs6$ apart. Star 1 was Joined with TIC ID 60340880, star 2 was included in TIC-8 as TIC 502852502.}
    \label{fig:oldsplit}
\end{figure}

In many cases, a single star detection by {\it 2MASS}, represented by a single entry in TIC-7, has been resolved into two or more stars by {\it Gaia\/} DR2. They can be very different in nature, from much fainter background stars to an unresolved double star of nearly equal brightness. TIC-8  contains 12.8 million stars resolved to be more than one source by {\it Gaia\/} DR2. These stars have the disposition SPLIT in TIC-8 and have the properties of the brighter {\it Gaia\/} DR2 star (see Figure~\ref{fig:oldsplit}). Thus a SPLIT disposition should be interpreted as follows:
\begin{itemize}
    \item The {\it 2MASS} magnitudes most likely are not reliable and should only be used if the secondary star is much fainter.
    \item The coordinates, proper motions and {\it Gaia\/} magnitudes of the original TIC ID have been set to those of the brighter {\it Gaia\/} star.
    \item Stellar characteristics like effective temperature, etc., have been replaced by those of the brighter {\it Gaia\/} DR2 star. These characteristics have been overwritten by those from specially curated lists like the Cool Dwarf Catalog.
    \item The brighter {\it Gaia\/} DR2 source is selected as the match to the {\it 2MASS} source, which we can refer to as TIC ID~1.  The other (usually the fainter) {\it Gaia\/} DR2 sources, can be found by searching for {\tt duplicate\_id} = 1.
\end{itemize}

\begin{figure}[!ht]
    \centering
    \includegraphics[width=0.225\linewidth]{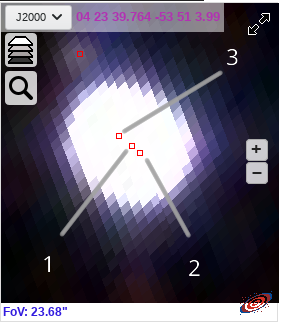}
    \caption{{\it 2MASS} color image of a newly found Split. One object was listed in TIC-7 (TIC 219253008, star 1). Star 2 is TIC 685340264, star 3 is TIC 685340263. Stars 1 and 2 have been Joined, and star 3 has no disposition, but its {\tt duplicate\_id} is set to 219253008 (the TIC ID of star 1).}
    \label{fig:newsplit}
\end{figure}

In a similar process as used for the Joins, we scanned the 2007 stars with TOIs for undetected Splits and found 47 unmatched stars from {\it 2MASS} that are candidates for Splits. We found potential Splits for 21 of them, using the same search radius as for Joins. The full list is in the Github repository and on MAST. For these new Splits we have 3 entries for 2 stars in the TIC-8; see Figure~\ref{fig:newsplit}. In this case entry 1 is the TIC-7 source which comes from {\it 2MASS}, star 2 is the brighter and star 3 the fainter star from {\it Gaia\/} DR2. It should be noted that in this case stars 2 and 3 have nearly the same brightness. We checked each case manually. 

The detailed procedure we developed for Splits is as follows:
\begin{enumerate}
    \item Coordinates are always copied from TIC ID 2 to TIC ID 1.
    \item Proper motions are copied from TIC ID 2 to TIC ID 1 if available from {\it Gaia\/} DR2.
    \item If star 1 is a member of a specially curated list like the Cool Dwarf Catalog, we simply adopt the stellar parameters from that list.
    \item If the {\it Gaia\/} DR2 quality flags in TIC-8 indicate that the properties derived from Gaia are acceptable (column {\tt gaiaqflag} is 1) we copy those from TIC ID~2 to TIC ID~1.
    \item We flag TIC ID~2 as DUPLICATE, to prevent it from being taken into account.
    \item We set the disposition of TIC ID~1 to SPLIT.
    \item For TIC ID~3 and any further stars we set their {\tt duplicate\_id} to TIC ID~1, but leave their disposition untouched.
\end{enumerate}

Any program ignoring stars with a DUPLICATE disposition will now find 2 valid entries for 2 physical stars.

\subsection{Correcting Flux Contamination Reported in the CTL}

Associated with the TIC is the Candidate Target List (CTL) as described in Section 3 of \citet{Stassun:2018tic} and updated in Section 3 of \citet{Stassun:2019}, which contains a ranked set of stars selected for TESS observations, optimized for transit detection.  For all stars in the CTL we calculate a rough estimate of the TESS flux contamination, using a number of simple assumptions about the TESS aperture and PSF.  That estimate is described in section 3.3.3 of \citet{Stassun:2018tic}.  
Here we report two corrections to that subsection.  

First, towards the end of that section just before an equation is provided for $N_{pix}$, there is a sentence that reads, ``We calculate a given star’s PSF using the formula below, requiring the PSF to be no smaller than one pixel and limiting the brightest stars to have a PSF size equal to that of a $T=4$ star."  In that sentence, the three mentions of ``PSF" should instead be ``aperture."

The second issue is that at the end of that section, step 3 provides the formula for the standard deviation of the 2-D Gaussian used to model the TESS PSF.  That formula is expressed in terms of the FWHM of the PSF, but the assumed value of the FWHM is not provided.  That number is $1.88 \times 20\farcs25$.

It is also the case that a much more accurate estimate of flux contamination is produced by the TESS SPOC pipeline for all stars observed at 2-min cadence.  That estimate is used by the SPOC pipeline to design the optimal aperture for that target, and the TESS data products use that estimate to calculate the CROWDSAP parameter included in TESS TPF file headers.

\section{All-sky search for phantoms (TIC 8.2)}\label{sec:allsky}

While manually checking TOIs, many of them after the TESS Follow-up Observing Program (TFOP) found discrepancies between TIC 8 and their observation, we faced the choice of not updating at all anymore, to continuously update TOIs as they are found or to try to find a general algorithm identifying artifacts, joins and splits all-sky. Given that the extended TESS mission includes a lot more targets from the Guest Observer Program, we decided to look for an all-sky solution.

\subsection{All-sky artifacts}\label{sec:allsky_artifacts}

We extended the artifact search described in \ref{sec:artifacts} to every {\it 2MASS} source with $T < 13$ where T is the TESS magnitude. Trying to use fainter limits produced more and more false positives; legit faint stars where the {\it 2MASS} pipeline could not produce good magnitudes for 2 of the bands. Given that $T < 13$ also is the limiting magnitude for the Candidate Target List (CTL) unless the star is provided by a specially curated list, we kept 13 as limit. The search produced $\approx 850000$ new artifacts.

\begin{figure}[!ht]
    \centering
    \includegraphics[width=0.4\linewidth]{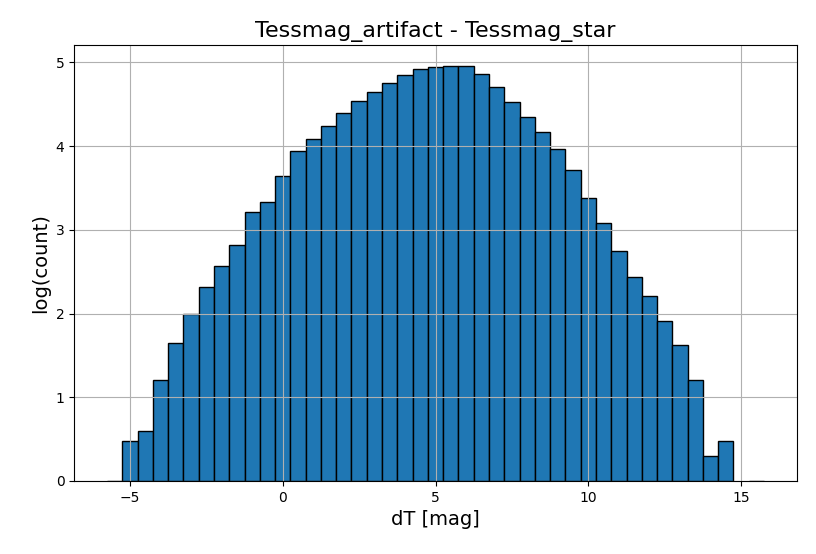}
    \includegraphics[width=0.225\linewidth]{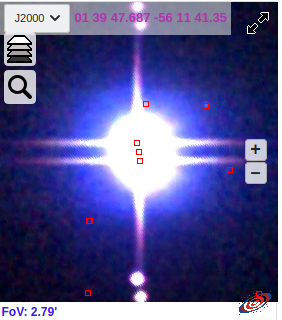}
    \caption{Left: histogram of TESS magnitude difference between artifact and the star it was found. Right: p Eridanus (TIC 231005906) as an example for a close double where the {\it 2MASS} pipeline falsely identified a third star (center marking) being brighter than the real components.}
    \label{fig:allsky_artifacts}
\end{figure}

Figure \ref{fig:allsky_artifacts} shows a histogram of the magnitude difference $T_{artifact} - T_{star}$ for all artifacts and the stars they have been found with. The histogram peaks at 5 and most of the artifacts at least 3 magnitudes fainter than their "host star". Surprisingly we found about $10000$ artifacts being brighter than the real star they surround. We identified 3 possible reasons for this:

\begin{enumerate}
    \item artifacts have at least 2 of the $J, H, K_{s}$ magnitudes with a U (worst) quality flag from 2MASS, thus the computed TESS magnitude can be off and the artifact is brighter in the computed TESS magnitude
    \item the artifact belongs to a bright star but happens to be found first with a fainter one
    \item several stars contribute to the artifact which then is brighter than any of the real sources.
\end{enumerate}

The right panel in Figure \ref{fig:allsky_artifacts} illustrates the third effect. It shows the {\it 2MASS} image for the known double p Eridanus, the red markings show the position of {\it 2MASS} sources. The center marking is that of an artifact most likely constructed by the {\it 2MASS} pipeline summing the flux of the components at this place. It should be noted that this gets quite common in the galactic plane.

It should be noted that this algorithm has it weaknesses, notably:

\begin{enumerate}
    \item limited by TESS magnitude, we can not positively exclude artifacts being around very red stars fainter than 13,
    \item we do not find all artifacts around very bright stars like Rigel or Betelgeuse because these can stretch for more than 1 degree on the sky.
\end{enumerate}

For any Guest Observer planning to propose observations, {\tt tic\_inspect} described in section \ref{sec:code} is a useful tool in case of doubt.

\subsection{All-sky joins and splits}\label{sec:allsky_joins}

Joins and splits can be identified by the same search algorithm:

\begin{verbatim}
    for every 2MASS-star without a Gaia ID and not being an artifact
        seachrad = 5 asec
        if tessmag >= 14, 15, 16
            searchrad = searchrad - 1, 2, 3  
            # means tessmag 16.5 has searchrad 2 asec
        candidates = all stars within searchrad with Gaia ID and without 2MASS ID
        if number of candidates == 1
            flag as join
        else
            if number of candidates > 5 and abs(galactic longitude) > 7.5
                skip because most likely a cluster
            else
                flag as split
\end{verbatim}

First of all only {\it 2MASS} without any counterpart from {\it Gaia} DR2 can be candidates for a join or a split. For these we define a search radius of 5 arc-seconds and adjust it in steps of -1 arc-seconds for stars fainter than TESS magnitude 14, 15 and 16. The default of 5 arc-seconds is roughly derived from the {\it 2MASS} PSF and from the experience we gained with manually checking for joins and splits. Going with a fixed value for the search radius turned out to be impractical because it produced many splits for very faint stars. Thus we looked at the {\it 2MASS} images for fainter TIC stars and found the limits above to be the most practical.

We then check the the number of candidates the search returns. If just one candidate is returned, we have a join, else we have a split. There is a chance however, that we end up in a very dense region outside of the galactic plane, such as Globular Clusters. In these regions we would end up with a mess of potential joins and splits, thus we try exclude them by skipping stars with more than 5 split candidates within the search-radius. This will not take out the outskirts of clusters, but the core regions. Guest Observers specifically interested in these regions are encouraged to use and adapt the {\tt tic\_inspect} script.

For this algorithm to work we had to break up the sky into search regions, because not all of the TIC 8 stars can be held in memory. This allows the same {\it Gaia} DR2 star being associated to 2 or more splits if it is located close to the border of 2 or more search regions. We accepted these cases at first and eliminated them in a correction run granting the uniqueness of any {\it Gaia} ID. Only the join or split with the closest distance on the sky remained.

\begin{figure}[!ht]
    \centering
    \includegraphics[width=0.36\linewidth]{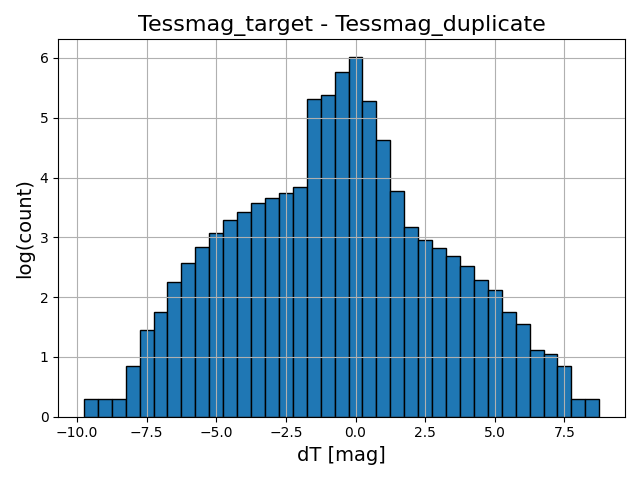}
    \includegraphics[width=0.225\linewidth]{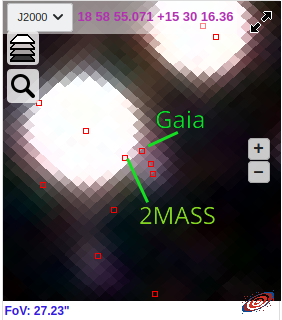}
    \includegraphics[width=0.225\linewidth]{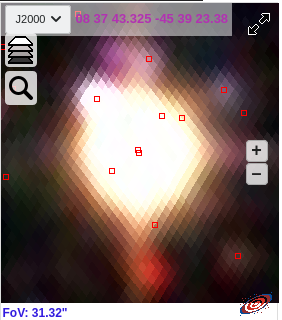}
    \caption{Left: histogram of TESS magnitude difference between join candidates and the star it was found with. The histogram centers around 0 as expected, but there are extreme left and right wings. Center: example from the extreme left end of the histogram (TIC 340546381). The {\it 2MASS} pipeline picks up a lot of flux from the bright star to the upper left, while {\it Gaia} with its higher resolution does not. Right: example from the extreme right end (the 2 close marks in the center). Here we got the TESS magnitude by an offset from {\it Gaia} DR1 which turned out to be much too faint.}
    \label{fig:allsky_joins}
\end{figure}

We found $3180382$ joins and $7948653$ splits. $818592$ joins turned out to be stars sitting right on top of known galaxies (TICID $> 10$ billion). In these cases we did not copy stellar characteristics, but simply flagged the star as duplicate. For the remainder of this text joins mean stellar joins where both object are stars.

Figure \ref{fig:allsky_joins} shows the magnitude differences between the join candidates in the left panel. The target is the {\it 2MASS} star, the duplicate the {\it Gaia} DR2 star it is going to be joined with. As expected the histogram centers around 0.0, but it shows extreme wings. The center panel shows an example from the extreme left end. In this case {\it 2MASS} picked up a lot of flux from the bright star to the upper left. {\it Gaia} DR2 with its better resolution did not, thus the {\it 2MASS} target is much brighter than the {\it Gaia} duplicate. The right panel is an example from the far right end of the histogram, the two markings close together at the center of the image are the ones in question. In this case we got the TESS magnitude of the target by an offset from the {\it Gaia} DR1 magnitude which turned out to be too faint.

The results for splits are similar to those for joins, except that magnitude differences are harder to interpret and use a quality check, because there are at least 3 stars involved now. Splits with only much fainter background stars around them will nearly keep their magnitudes while double-stars with similar magnitudes will get fainter, because 1 {\it 2MASS} entry is split into 2 {\it Gaia} stars. As mentioned in Section \ref{sec:howto} the stellar characteristics of the brighter {\it Gaia} star will be copied to the {\it 2MASS} target, thus the TESS magnitude of the {\it 2MASS} target is corrected. The brighter {\it Gaia} star then is flagged as duplicate and should not be taken into account anymore.

\subsection{Quality check}

\begin{figure}[!ht]
    \centering
    \includegraphics[width=0.8\linewidth]{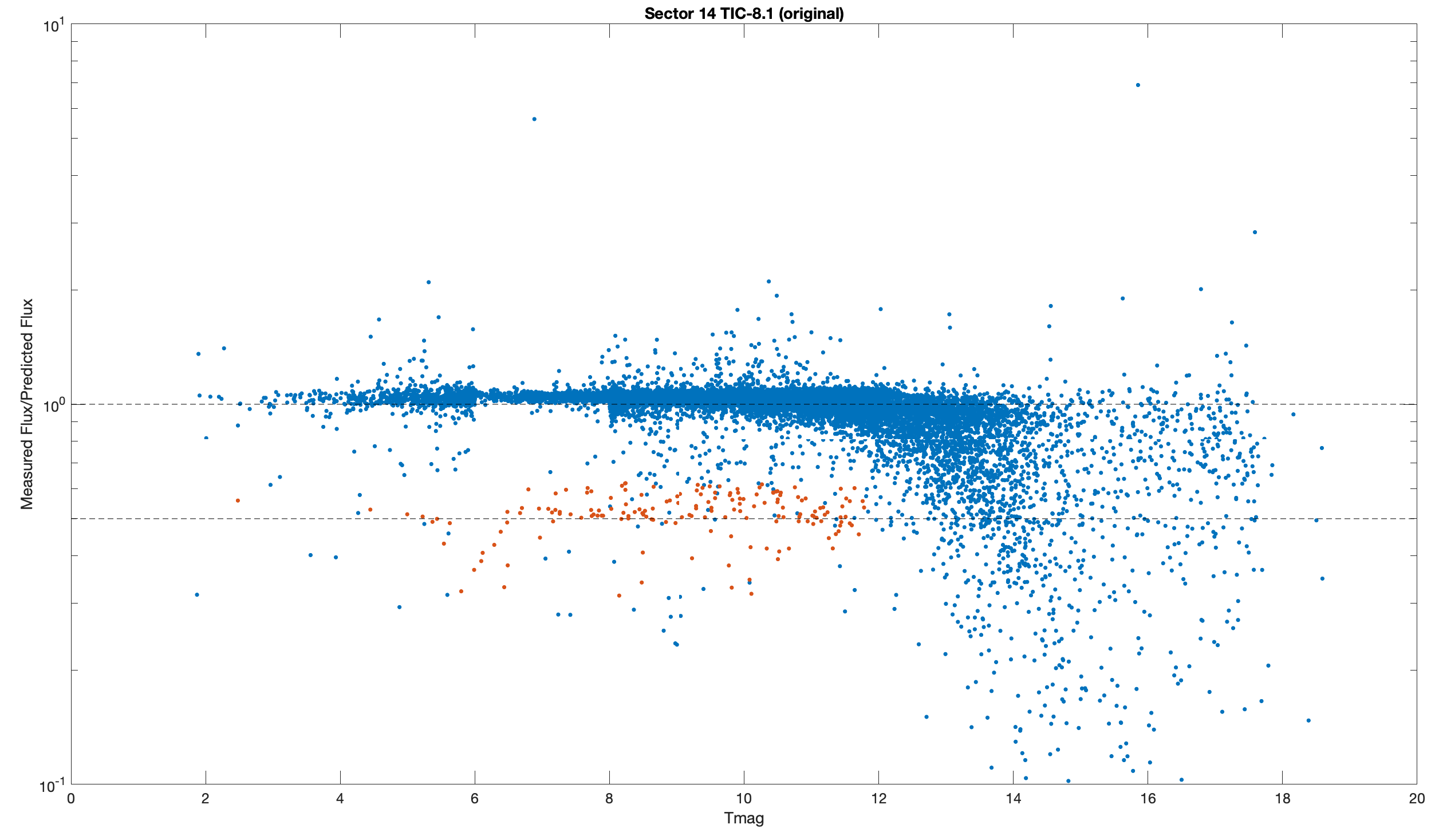}
    \includegraphics[width=0.8\linewidth]{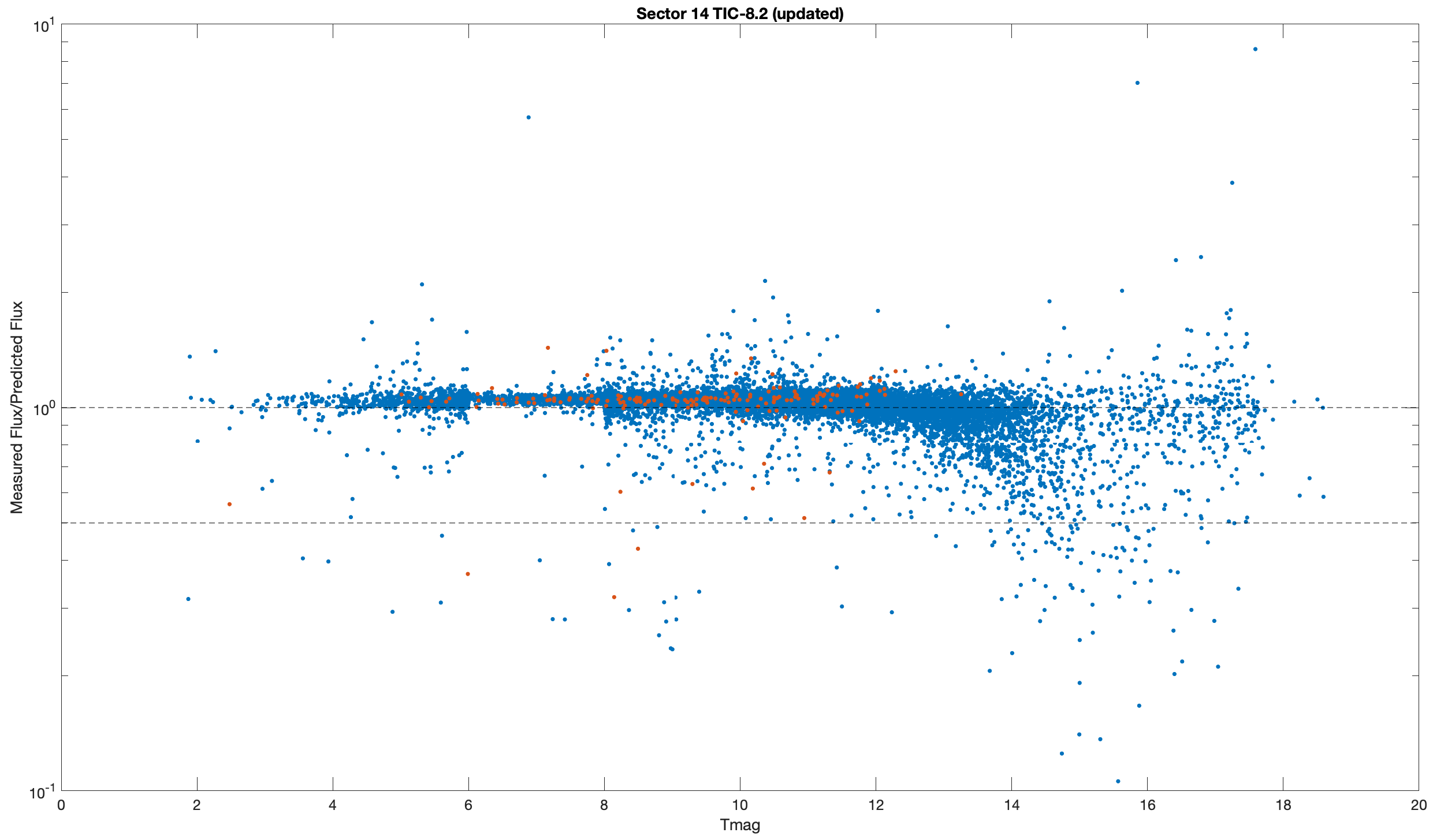}
    \caption{Flux ratios measured / expected flux. Top: result using TIC 8.1, stars affected by corrections in TIC 8.2 are marked in orange. Bottom: ratios after reprocessing all 2-minute targets with results from this paper, orange marks identify the same stars as in the top panel. The narrower distribution of faint stars is due to improvements in the SPOC processing pipeline.}
    \label{fig:sect14}
\end{figure}

TESS Sector 14 was the first sector to be reduced using TIC 8. It turned out that for $\approx 1 \%$ of the $20000$ 2-minute-targets the measured flux deviated from the expected flux from TESS magnitudes by more than a factor 0.3, with the measured flux being less than expected. In parts this is expected, say for variable stars. We provided TESS Science Processing Operation Center (SPOC) with our results for sector 14 and they re-processed the sector. The upper panel of Figure \ref{fig:sect14} shows the ratios of measured to expected flux with stars affected by our corrections marked in orange. The lower panel shows the same after reprocessing, moving most of the marked stars to a flux ratio close to 1. We inspected the few remaining orange dots manually and found that all except 2 of them are known variable stars.

\section{Discussion: Data Access and Future Reductions}\label{sec:discussion}

We are well aware of the fact that our all-sky search and correction improves the TIC, but is not perfect. We can not grant that no further artifacts, joins or splits are found in the future, but we expect to have driven the probability from $\approx 1.1 \%$ down to $ < \approx 0.1 \%$. The results from sector 14 support this claim. Artifacts, joins and splits found from manually inspected TOIs (Section \ref{sec:results}) have been made available at MAST. All changes from the all-sky search have been sent to MAST and will be integrated into the official TIC shortly, thus comprising TIC version 8.2. 

In addition, we provide the {\tt tic\_inspect} script (Appendix~\ref{sec:code}) which scans the neighborhood of stars of interest for Artifacts, Joins, and Splits and thus helps the community to correct for them in their research. At this point we do not re-compute the contamination ratios and priorities given in TIC 8.1 and the Candidate Target List (CTL). Reasons:

\begin{enumerate}
    \item our contamination rate is an estimate assuming a Gaussian PSF, POC uses its own contamination algorithm taking into account the position on the camera and other details
    \item we computed contamination rates only for CTL stars while more and more targets are selected for a short cadence which are not in the CTL
    \item it is too time-consuming to compute contamination ratios for all TIC star and then to update the whole TIC.
\end{enumerate}

Instead we provide the {\tt tic\_contam} script, allowing to compute the contamination ratio for any given target (see Appendix \ref{sec:code_contam}).

\section{Summary}\label{sec:summary} 

The TESS Input Catalog \citep[TIC;][]{Stassun:2019} is a comprehensive collection of $1.73\times10^9$ sources on the sky. 
Certain types of spurious entries in such a large catalog representing the federation of multiple smaller catalogs are unavoidable, and while representing a very small fraction of the overall TIC, they can lead to incorrectly estimated transiting planet radii and wasted followup effort. 

We have defined three types of spurious entries in the TIC: Artifacts, Joins, and Splits. 
We generally refer to these problems as ``phantoms", which refer to objects listed in TIC-8 that are not real or where the number of objects associated with a given TIC ID or set of IDs is incorrect. 
We have developed an approach by which these phantoms can be identified and a methodology for correcting the TIC-estimated flux contamination as appropriate, which in turn permits a corrected planet radius via the corrected transit depth. 

Finally, we have supplied corrections for all of the known phantoms at the time of this writing to the MAST database service. We also make available a code, {\tt tic\_inspect} and {\tt tic\_contam}, by which any other phantoms may be identified (Appendix~\ref{sec:code}) and contamination ratios for any object can be computed (Appendix~\ref{sec:code_contam}).

\acknowledgements
We gratefully acknowledge NASA grant 17-XRP17 2-0024 for partial support of the work to develop value-added tools for the community's use of the TESS Input Catalog and Candidate Target List. Screenshots of catalogs have been made using the VizieR catalogue access tool, CDS, Strasbourg, France (DOI : 10.26093/cds/vizier). The original description of the VizieR service was published in 2000, A\&AS 143, 23, \citep{vizier:2000}.

\appendix

\section{Sample List of Manually Inspected Phantoms}\label{sec:appendix}

Since the release of TIC-8, there has been an effort to identify these phenomena, and preliminary lists of Joins, Splits, and Artifacts have been assembled and posted to the TIC/CTL Data Release Notes website on MAST\footnote{\url{https://outerspace.stsci.edu/display/TESS/TIC+v8+and+CTL+v8.xx+Data+Release+Notes}}.  Below we display representative portions of tables of manually confirmed Joins, Splits, and Artifacts. The full tables are also available electronically at the Github repository for the {\tt tic\_inspect} code (see Appendix~\ref{sec:code}). 

\begin{verbatim}
Artifacts
TIC ID     ! 2MASS ID        ! comments
----------+------------------+------------
269701145 ! 20080397+6651023 ! near TOI 1339
 27462105 ! 14274226+4157123 ! near 158025009 ! 14274177+4157124 !
178819687 ! 12575199-3945275 !
178819685 ! 12575242-3945196 !
 94986318 ! 05272486-1416301 !

Joins
 TOI    ! TIC ID 1   ! TIC ID 2  ! comments
--------+-----------+------------+----------------------------------------------
 454.01 ! 153077621 !  651051509 ! TIC ID 1 is a cool dwarf
 573.01 ! 296780789 !  876200724 ! TIC ID 1 is a cool dwarf
 722.01 !  38509907 !  684936227 ! TIC ID 1 has a spectroscopic Teff
1421.01 ! 329277372 ! 2019613336 ! Gaia photometry unreliable, RUWE = 41.2 (not ok)
        ! 430528563 ! 2014876481 ! this is Tycho-2 3969-00744-1

Splits
 TOI    ! TIC ID 1  ! TIC ID 2   ! TIC ID 3   ! comments
--------+-----------+------------+------------+-----------------------------------
        !  92359852 !  651667036 !  651667038 ! RUWE < 1.1 for TIC ID 1 and 2 (ok)
1606.01 ! 125735470 !  640763859 !  640763860 ! TIC ID 1 and 2 have similar properties
 268.01 ! 219253008 !  685340264 !  685340263 ! TIC ID 1 and 2 have similar properties
1014.01 !  96246348 !  772008799 !  772008800 ! TIC ID 2 is brighter, but too hot for properties
1044.01 ! 463402815 !  854828119 !  854828120 ! flags for TIC ID 2 not ok, but RUWE = 0.93

\end{verbatim}

\section{The tic\_inspect script}\label{sec:code}

In order to allow TIC-8 users to check interesting targets we wrote a short Python3 script named tic\_inspect.py. The script, a sample input file, and resulting tables from scanning the TOIs are publicly available on Github as \url{https://github.com/mpaegert/tic_inspect}. This script is written in Python 3 and makes use of {\tt astropy} and {\tt astroquery.mast}. It has been tested on different Linux distributions with and without Anaconda. On Windows~10 it can be invoked via {\tt python tic\_inspect.py}.

Running the script without any parameters will print a short help about its command-line parameters. The script needs either the TIC ID or an input list of TIC IDs (command-line flags {\tt --ticid} or {\tt --ticidfile}). The file should contain a comma-separated list of TIC IDs to inspect - in one or more lines. For each TIC ID the script will query the TIC-8 table on MAST for a cone around that star, then propagate proper motions from J2000.0 to a given target epoch (default: J2019.5, thus between year 1 and 2 of the TESS mission). The resulting list of nearby stars is updated with the proper-motion corrected coordinates, the radial distances are re-computed using these coordinates, and the table is finally sorted by the updated radial distances. This updated table is then scanned for Artifacts, Splits and Joins. The result is written in formatted form to a text-file (default: {\tt report.txt}) and as csv table to a second file ({\tt phantoms.csv}). 

By default, tic\_inspect runs a cone-search $20\farcs25$ around the target star. This cone is searched for Artifacts, the result printed and the identified Artifacts deleted from the table in order to take them out of the search for Joins and Splits. The table is then cut down to the search-radius for Joins and Splits (default: $5\farcs0$) and the rest searched for stars which have a {\it Gaia\/} DR2 ID, but no {\it 2MASS} ID. If one star is found, the target star is seen as candidate for a Join. If more stars are found we treat the target star as candidate for a Split.

It should be noted that the search radius for Joins and Splits (parameter {\tt --joinrad}) plays an important role. The greater the search radius, the fewer Joins and more Splits the script will identify. 

Using the report produced by tic\_inspect and the resulting lists of nearby stars, it is possible for a user to at least manually trace the steps we take for detecting and dealing with Artifacts, Joins and Splits.

\section{The tic\_contam script}\label{sec:code_contam}

This script has the same requirements and works much like {\tt tic\_inspect}. Calling it without any parameter produces a short help, calling it with a TICID or file with TICIDs to process will compute the ration of target flux to flux by surrounding contaminating stars. Any additional artifacts or duplicates found by {\tt tic\_inspect} should be given in an file with TICIDs to exclude and the script should then be called with the {\tt --ticexfile} parameter.

All TESS-specific settings are defined as global parameters at the beginning of the script, starting with {\tt tess\_pixel}. Changing the constants allows to adapt the script to other telescopes than TESS.

\bibliographystyle{aasjournal}
\bibliography{references}

\end{document}